# A floating top-electrode electrowetting-on-dielectric system†

Hanbin Ma, [*ab] Siyi Hu, [a] Yuhan Jie,[b] Kai Jin[a] and Yang Su[*ab]

Herein, we describe a novel device configuration for a double-plate electrowetting-on-dielectric system with a floating top-electrode. As a conventional double-plate EWOD device requires a grounded electrode on the top-plate, it will cause additional fabrication complicity and cost during the encapsulation process. In this work, we found that by carefully designing the electrode arrangement and configuring the driving electronic circuit, the droplet driving force can be maintained with a floating electrode on the top-plate. This can provide the possibilities to integrate additional electrical or electrochemical sensing functions on the top-plate. We use both finite element analysis and the fabricated system to validate the theory, and the results indicate that floating top-electrode EWOD systems are highly reliable and reproducible once the design considerations are fully met.

## 1 Introduction

Electrowetting-on-dielectric (EWOD) is a branch of microfluidic devices which has been rapidly developed in the last two decades.[1,2] By feeding sequenced voltage signals to an on-chip electrode array, the digital microfluidic chip could enable complex manipulation of each discrete droplet. The platform could support a wide range of applications, which provides the possibility to enable true lab-on-a-chips.[3–5] One main advantage of EWOD is that no prefabricated micro-channels are required, the liquid droplets are manipulated in a two-dimensional region by electrical signal-induced surface tension changes. However, the major drawback is that the droplet can only be controlled on the surface of the area where electrodes are physically located. A large number of electrodes or an electrode array is highly demanded where a complex bio-medical procedure which requires simultaneous control of multiple liquid samples is conducted. The driving electronics also need to be fast and reliable to drive the electrodes.

Thanks to the fast development of consumer electronics, the nowadays semiconductor technologies could fulfil the EWOD's requirement for sequenced voltage control easily. Pioneer researchers have also published the details of their systems, and the open-source platforms (e.g. DropBot and Opendrop) have promoted the development of EWOD greatly.[6,7] In general, an EWOD system can be depicted as the block diagram shown in Fig. 1(a). A software is used to communicate with a control unit (e.g. an ARM-based microcontroller in Opendrop). The control unit will then instruct a switching unit to feed power-source-generated voltage signals to the microfluidic chip. The selection of the switching unit is the key to the system design, which may influence the overall system performance greatly. Solutions, e.g. discrete optical relay,[2,6] commercial integrated switch box (with relays),[8,9] solid-state high-voltage multiplexer,[7,10] thin-film electronics-based pixel circuit[11,12] and others[13,14] have been reported previously.

For the EWOD microfluidic chip arrangement, there are two technical routes available: single-plate and double-plate. For single-plate setup, the electrodes sit on a substrate (the bottom plate). A layer of dielectric and a layer of hydrophobic coating are on the top of the electrodes, and the hydrophobic coating serves as the interface between the plate and the target sample droplet(s). The double-plate arrangement uses one additional substrate to cover the top of the sample droplet(s) to create a sandwich structure. The single-plate devices and the double-plate ones could be driven by the same electronic system. Although single-plate devices have a simple structure, complex operation e.g. splitting droplets can only be realized by a double-plate architecture. Double-plate devices can precisely define the thickness of a sample droplet by controlling the gap between the two plates. Meanwhile, the direct electrical field in the double-plate setup can also provide a more effective electrowetting force to overcome the surface tension, while the electric potential is distributed as a fringe field in a single-plate device. In addition, the two plates provide the capability to introduce liquid medium (rather than air) to further optimize the surface tension at the droplet–medium interface.

In conventional double-plate EWOD system, the top-plate is a glass substrate with a layer of transparent conductive oxide

[a]CAS Key Laboratory of Bio-medical Diagnostics, Suzhou Institute of Biomedical Engineering and Technology, Chinese Academy of Science, No. 88 Keling Road, Suzhou, Jiangsu Province, 215163, P. R. China. E-mail: mahb@sibet.ac.cn
[b]ACXEL Tech Ltd, Unit 184 Cambridge Science Park, Cambridge, CB4 0GA, UK. E-mail: yang.su@acxel.com
† Electronic supplementary information (ESI) available. See DOI: 10.1039/c9ra09491a

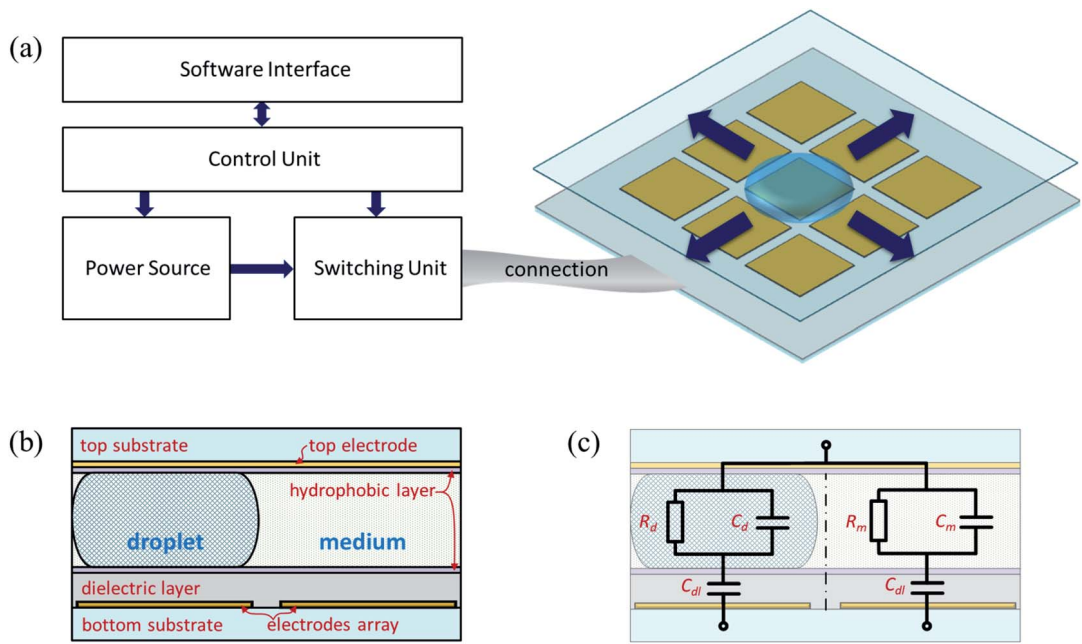

Fig. 1 Common configuration of EWOD systems. (a) Block diagram of the electrical driving system, the connection and the EWOD device with an electrode array. (b) Cross-section of a conventional double-plate EWOD device. (c) An equivalent circuit of a double-plate EWOD device.

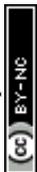

(*i.e.* indium-tin oxide) to provide a clear view of the sample droplet(s). A thin layer of non-insulating hydrophobic coating is then added to the surface to ensure an electrical connection with the sample droplet(s). In most of the reported system, the conducting layer of the top-plate needs to be always physically connected to the grounded point on the control electronic circuit. Swyer *et al.*[15] have recently reported an EWOD system where a dielectric layer is applied to the conductive top plate without affecting the droplet manipulation. Their work is considered as a transit between single-plate and double-plate devices. However, the electrode on the top-plate is still connected to an electrical grounded point. Here, we take one step further, for the first time, we report a double-plate EWOD system with a floating top-electrode. From both theoretical simulations and real device validation tests, we confirm that once the number or the area of grounded bottom electrodes is large enough, the double-plate device can work effectively with a floating top-electrode. The reported new setup can reduce the encapsulation cost of a double-plate EWOD device significantly. The floating top-electrode can be electronically isolated from the EWOD driving system to realize additional electrical or electrochemical sensing functions. For example, we have demonstrated an impedance-based sensing system for monitoring multiple connections between EWOD electrodes and driving electronics.[16] The floating top-electrode EWOD system design can be considered as a new class of device configuration to the conventional single-plate and double-plate setups.

## 2 Theory

Fig. 1(b) shows a conventional double-plate EWOD microfluidic device configuration. A bottom substrate hosts an electrode array (only two bottom electrodes are shown in the figure), a dielectric layer and a hydrophobic layer. The top substrate has a continuous electrode with a hydrophobic coating which is faced to the bottom. The gap between the two substrates is normally defined by a spacer which is not shown in the figure. The sample droplet sits between the two plates (normally on top of an electrode) with the surrounding medium. Fig. 1(c) illustrates the equivalent circuit for the double-plate device. $C_{dl}$ is the capacitance of the dielectric layer, $R_d$ and $C_d$ is the impedance of the sample droplet, and $R_m$ and $C_m$ is the impedance of the medium. Since the thickness of the hydrophobic layer is so thin, the impedance of these layers is neglected. For a simplified two-electrode system, there is three terminals in the equivalent circuit: two bottom electrodes and a top-electrode.

Fig. 2(a) shows an advanced model for a moving droplet between two adjacent electrodes, which was introduced previously.[15,17] The length of the droplet and the pitch of the electrodes are identical as $L$, the droplet moving distance from the original electrode ($E_1$) to the destination electrode ($E_2$) is $x$, the thickness of the bottom dielectric is $t$, and the gap between the two plates is $g$. Fig. 2(b) is a modified equivalent circuit based on the definitions in Fig. 2(a). To simplify the calculation, the resistive elements are neglected here. The three terminals are indicated as hollow circles. Fig. 2(c) shows the circuit connection of a conventional double-plate EWOD device. The top-electrode is constantly connected to ground, and the bottom electrodes are connected to a switching unit which can be turned ON or OFF as required. Once $E_2$ is ON, the sample droplet will start moving from $E_1$ (at OFF state) to the ON-state electrode $E_2$. Depending on the design of the switching unit, the OFF-state electrode $E_1$ can be grounded, floating or high-$Z$. Fig. 2(d) shows an alternative connection where the top-electrode is floating (not connected to the electronic driver). Fig. 2(e) and (f) are the equivalent circuits of the above two connection methods.

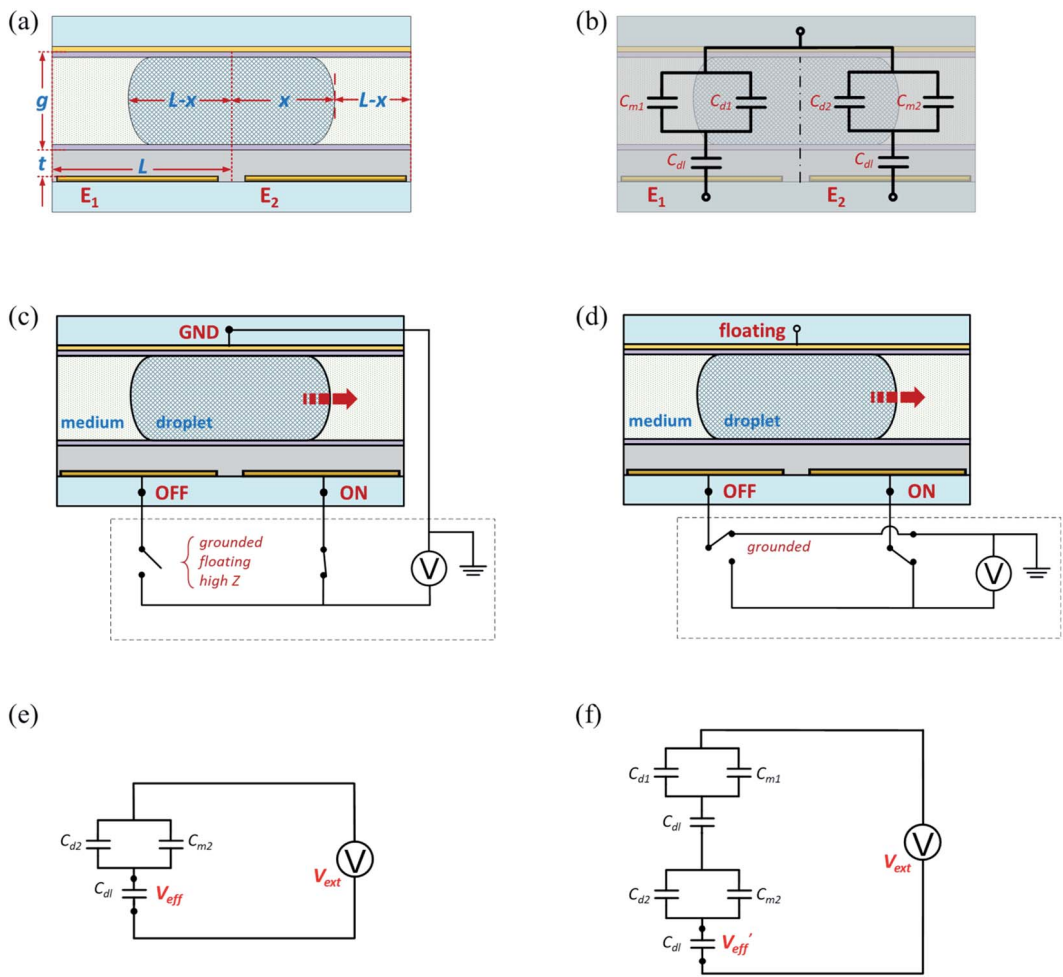

Fig. 2 Comparison between conventional double-plate EWOD device and floating top-electrode EWOD for moving one sample droplet from one electrode to another adjacent electrode. (a) Dimensions of a moving droplet over two adjacent electrodes. (b) An advanced equivalent circuit of a moving droplet, where the resistive elements are neglected. (c) Circuit connection for a conventional double-plate EWOD device. (d) Circuit connection for a floating top-electrode EWOD device. (e) Advanced equivalent circuit model for conventional grounded top-electrode connection as shown in (c). (f) Advanced equivalent circuit model for floating top-electrode connection as shown in (d).

For EWOD systems, the droplet driving force is introduced by the voltage-induced contact angle changes, and the misbalanced contact angles at different directions will then drive the droplet to move.[18] It has been proven that the voltage drop on the dielectric layer is the key to affect the interface surface tension, which leads to a contact angle change.[8,19,20] Therefore, the voltage drop over the dielectric layer on the top of the driving electrode (the ON-state electrode which is $E_2$ in this case), is the effective voltage, $V_{eff}$. Once the effective voltage (at the droplet–medium interface) is greater than the threshold voltage ($V_{th}$), the droplet can be moved from one electrode to another. In the equivalent circuit shown in Fig. 2(b), the capacitance elements can be calculated by the following equations:

$$C_{m1} = \frac{\varepsilon_m \varepsilon_0 yx}{g}; \quad C_{m2} = \frac{\varepsilon_m \varepsilon_0 y(L-x)}{g}; \quad (1)$$

$$C_{d1} = \frac{\varepsilon_d \varepsilon_0 y(L-x)}{g}; \quad C_{m2} = \frac{\varepsilon_d \varepsilon_0 yx}{g}; \quad (2)$$

$$C_{dl} = \frac{\varepsilon_m \varepsilon_0 yL}{t}; \quad (3)$$

where $y$ is the width of the droplet and the electrodes.

In a conventional connection method based on the models in Fig. 2(c) and (e), the $V_{eff}$ under an external driving voltage $V_{ext}$ can be calculated as:

$$V_{eff} = V_{ext} \frac{\frac{1}{C_{dl}}}{\frac{1}{C_{d2}+C_{m2}}+\frac{1}{C_{dl}}} \quad (4)$$

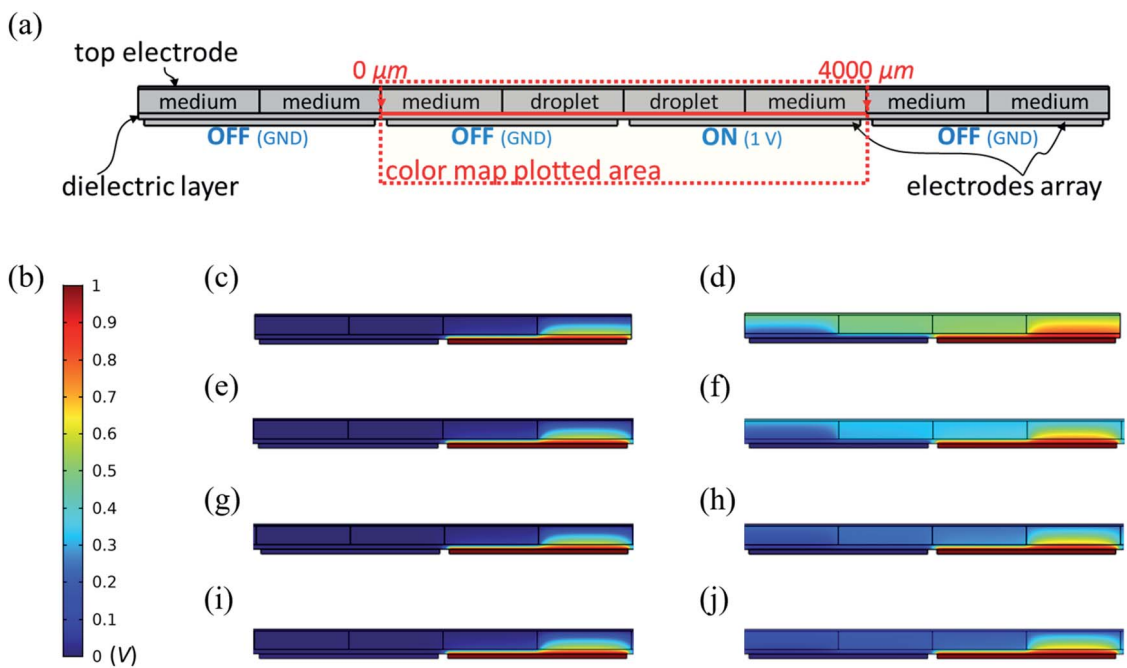

Fig. 3 Finite element analysis of EWOD device in COMSOL Multiphysics. (a) 2D model for a double-plate EWOD device, eight electrodes are created in the model, and only four are presented in this figure; one electrode is set at ON-state with 1 V, and the rest are set as OFF-state (grounded); a droplet sits in the middle and is surrounded by the medium; the length of the droplet is 2000 μm which is the same as the pitch of the electrodes; the thickness of dielectric layer is set as 50 μm, and the gap is 200 μm. (b) The scale bar of the electric potential color maps in (c)–(j). (c) Electric potential color map of a two-electrode system (1 ON-state and 1 OFF-state, $\kappa = 1$) with top-plate grounded. (d) Electric potential color map of a two-electrode system ($\kappa = 1$) with a floating top-electrode. (e) Electric potential color map of a four-electrode system (1 ON-state and 3 OFF-state, $\kappa = 3$) with top-plate grounded. (f) Electric potential color map of a four-electrode system ($\kappa = 3$) with a floating top-electrode. (g) Electric potential color map of a six-electrode system (1 ON-state and 5 OFF-state, $\kappa = 5$) with top-plate grounded. (h) Electric potential color map of a six-electrode system ($\kappa = 5$) with a floating top-electrode. (i) Electric potential color map of an eight-electrode system (1 ON-state and 7 OFF-state, $\kappa = 7$) with top-plate grounded. (j) Electric potential color map of an eight-electrode system ($\kappa = 7$) with a floating top-electrode.

For a floating top-electrode connection, assuming the OFF-state electrode $E_1$ is connected to the ground, an equivalent circuit can be extracted as shown in Fig. 2(f), and the $V'_{eff}$ can be calculated as:

$$V'_{eff} = V_{ext} \frac{\frac{1}{C_{dl}}}{\frac{1}{C_{d2} + C_{m2}} + \frac{1}{C_{dl}} + \frac{1}{C_{d1} + C_{m1}} + \frac{1}{C_{dl}}} \quad (5)$$

The above equation is based on a two-electrode system assumption, where there is only one OFF-state electrode and one ON-state electrode. However, in a realistic EWOD system, more electrodes are required, and the number of OFF electrodes is normally much bigger than that of ON electrodes. Here, we define the OFF-electrode number to ON-electrode number ratio is $\kappa$. When there are additional OFF-state electrodes available in the system with the only medium in the gap, the eqn (5) can be then re-written as:

$$V'_{eff} = V_{ext} \frac{\frac{1}{C_{dl}}}{\frac{1}{C_{d2} + C_{m2}} + \frac{1}{C_{dl}} + \frac{1}{\frac{C_{dl}(C_{d1} + C_{m1})}{C_{dl} + C_{d1} + C_{m1}} + \frac{\kappa C_{dl}(C_{m1} + C_{m2})}{C_{dl} + C_{m1} + C_{m2}}}} \quad (6)$$

For a double-plate EWOD system, if $V'_{eff}$ is greater than a contact angle hysteresis induced $V_{th}$, then it can be operated at a floating top-electrode mode. If $\kappa$ is big enough, the voltage between the grounded electrodes and floating top plate can be eventually neglected, where $V'_{eff} \approx V_{eff}$.

## 3 Experimental

### 3.1 EWOD system

A double-plate EWOD system was designed to validate the theory in this work. The bottom plate was a standard printed circuit board (PCB) with an array of electrodes. Conventional Saran wrap (50 μm-thick polyethylene, PE) was used as the dielectric layer. Cytop from AGC was used as the hydrophobic coating for both the bottom plate and the ITO-coated glass top plate. A 100 μm gap was realized by a PTFE spacer. 5 cSt silicon oil from Dow Corning was used as the medium, and DI water with food dye was used as sample droplets. High-voltage solid-state multiplexers (HV507 from microchips) were used to provide the sequenced voltage signals for the electrode array. The switching unit and a customized power source were controlled by a microcontroller (STM32H7 from STMicroelectronics). A customized software interface was also used in this work, which communicated with the microcontroller through a series bus.

### 3.2 Finite-element analysis

A two-dimensional finite element analysis for a double-plate EWOD device was performed by COMSOL Multiphysics. As shown in Fig. 3(a), the model consisted of an electrodes array, a dielectric layer, a top-electrode and a gap for droplet and medium. One electrode was set as ON-state at 1 V, and the rest were set as OFF-state at 0 V (grounded). The electrode pitch was 2000 μm, and a 100 μm space was set between the electrodes to avoid short-circuit. On the top of the bottom electrodes, a 50 μm-thick dielectric layer was placed. A gap between the top-electrode and the dielectric was 200 μm, and it was divided into a few identical rectangles with the length of 1000 μm. The two middle rectangles represent the sample droplet indicating that half of the droplet sited on the top of an OFF electrode and the other part sited on an ON electrode. The rest of the gap was filled with the medium. For electric potential simulation, only the targeted area is plotted, which is indicated by a red dashed square in Fig. 3(a). A red solid cutline is also used in the figure to show the dielectric interface potential from 0 μm to 4000 μm. Two-electrode, four-electrode, six-electrode and eight-electrode systems were simulated separately. For all simulations, only one electrode was set as ON-state while the others were all grounded. All dimensions and materials' properties were set based on our EWOD system described in Section 3.1.

### 3.3 Speed test for droplet manipulation

Droplet manipulation speed is a key parameter of digital microfluidics, and it is used in this work to characterize the performance of floating top-electrode EWOD system. We compared the speed of single droplet transportation using a 300 V DC signal. The electrode pattern is shown in Fig. 5(a), and two electrode arrays with different sizes were used for comparison ($n_1 \approx 1.39$ mm and $n_2 \approx 1.90$ mm). The surrounding area (with no electrode features) served as a common electrode which was connected to ground all the time. With the common grounded electrode, we maintained our system with a big effective $\kappa$ all the time. During the measurements, five linearly arranged electrodes (as indicated in Fig. 5(a)) were used to perform a continuous single-droplet movement. The movement step was precisely controlled by a timer defined in the MCU with a resolution of 0.01 s. We use the completion rate to quantitatively analyse the speed, which is defined by the successful moving electrode number over that of total linear electrode number, which is five in this case.

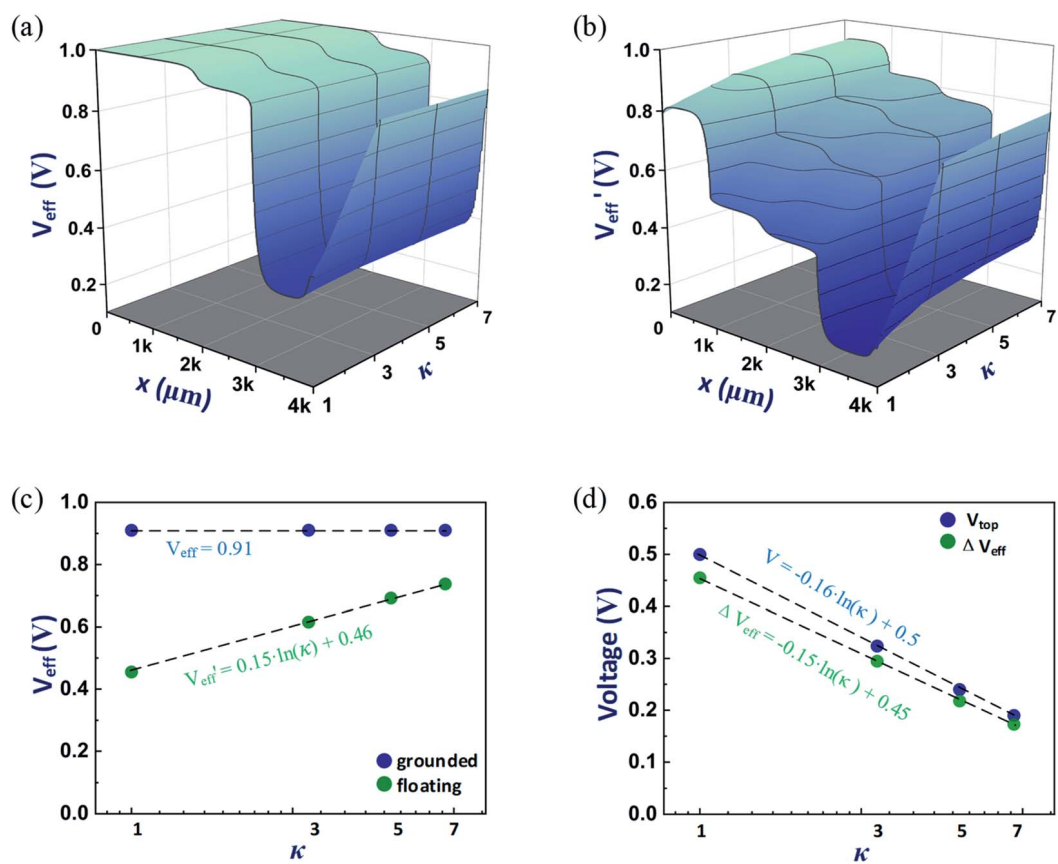

Fig. 4 Effective voltage comparison between grounded top-electrode group and floating top-electrode group. (a) 3D color map of EWOD effective voltage drop ($V_{eff}$) with grounded top-electrode with different $\kappa$. (b) 3D color map of EWOD effective voltage drop ($V'_{eff}$) with floating top-electrode with different $\kappa$. (c) Effective voltage at the sample droplet and surrounding medium interface (the cliff point at $x = 2950$ μm) of grounded top-electrode and floating top-electrode groups in semi-log scale. (d) Top-electrode voltage changes in floating top-electrode group, and the effective voltage differences ($\Delta V_{eff}$) at different $\kappa$.

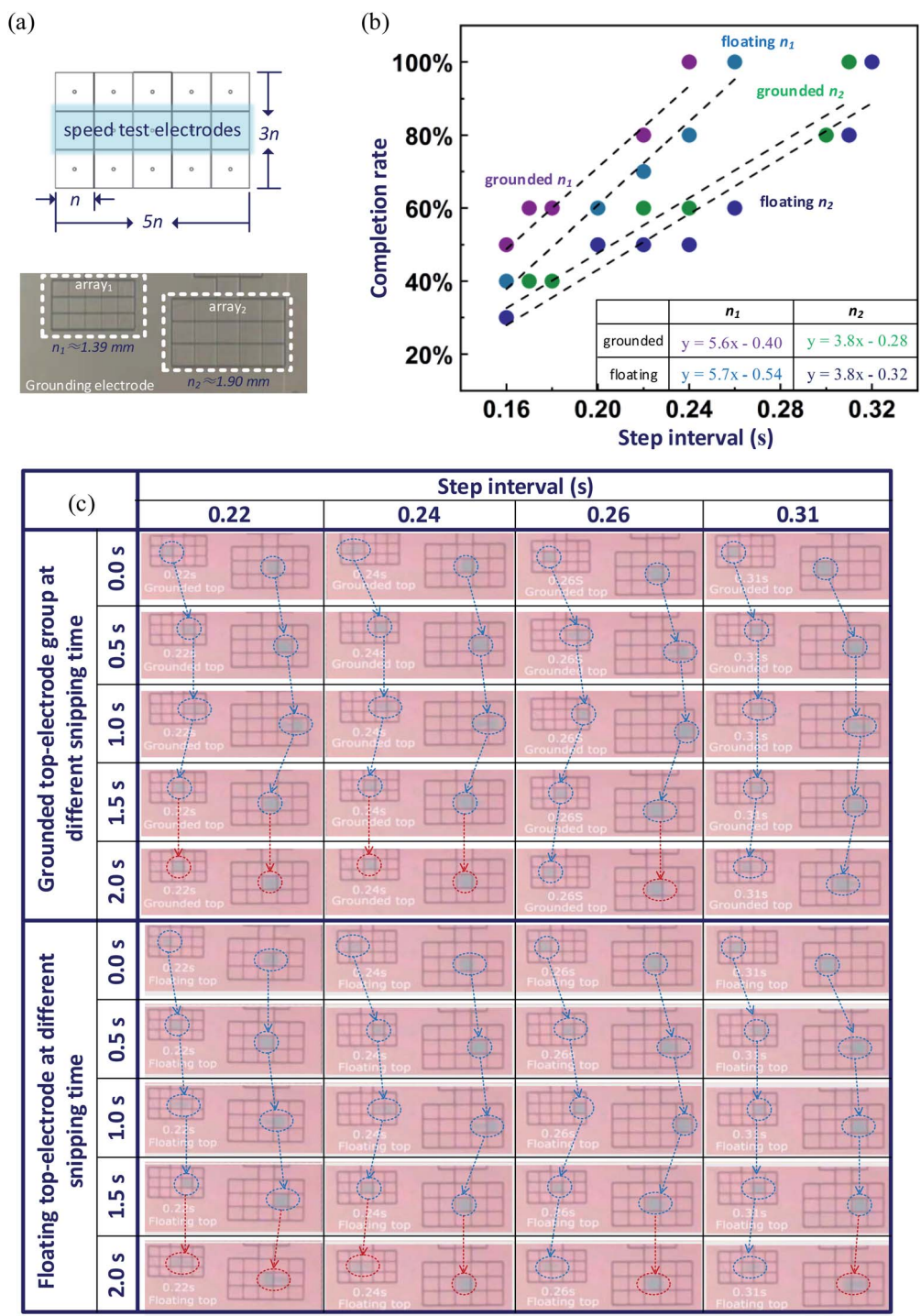

Fig. 5 Droplet movement speed test for grounded and floating top-electrode setups with two different-size electrode arrays. (a) Schematics for a 5 by 3 electrode array, and a picture for the fabricated electrodes (electrode length for $n_1$ is around 1.39 mm, for $n_2$ is around 1.90 mm). (b) Completion rate results with different step intervals for $n_1$ and $n_2$ with grounded and floating top electrodes; linear fitting equations are listed in the sub-table. (c) Snip pictures with four step intervals (0.22 s, 0.24 s, 0.26 s and 0.31 s) at 0.0 s, 0.5 s, 1.0 s and 2.0 s for both grounded top-electrode group and floating top-electrode group.

## 4 Results and discussion

### 4.1 Electric potential simulations

Fig. 3(b) is the scale-bar of the color-map simulations, a 1 V voltage stimulus was used here to simplify the calculation, and grounded electrodes were all set to 0 V. Fig. 3(c)–(j) show the simulations under different connection conditions, and with different effective electrode numbers. Four groups of simulations were conducted for $\kappa$ (OFF-electrode number to ON-electrode number ratio) = 1, 3, 5 and 7. Fig. 3(c)–(j) plot the potential distribution of the centre parts where the droplet movement happens, and the full simulated results are shown in the ESI.† In the plotted area, the left electrode is the droplet's original position and is grounded. The right electrode is the destination and is connected to the voltage stimulus. A cutline was also created to investigate the voltage drop along with the dielectric layer. As shown in Fig. 3(a), the cutline was set at the interface between the dielectric layer and the medium/droplet layer at x-direction from 0 μm to 4000 μm.

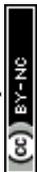

For the grounded top-electrode group (Fig. 3(c), (e), (g) and (f)), the electric potential distributions are identical for different electrode numbers. It is because a direct-gap capacitance is formed between the driving electrode and grounded top-electrode, the field distribution is identical regardless of the change of the electrode number. On the contrary, for the floating top-electrode groups (Fig. 3(d), (f), (h) and (j)), the field distributions change gradually as $\kappa$ increases. Fig. 4(a) and (b) show the effective voltage drop on the dielectric layer for two groups of simulations. Fig. 4(a) is the grounded top-electrode group, where the voltage drop profiles of all four simulations are identical. Fig. 4(b) is the floating-top-electrode group, the $V'_{\text{eff}}$ increase gradually as $\kappa$ increase. The cliff points for all simulations show at the interface between the droplet and medium (x-direction 2950 μm), where the droplet driving force is generated by the contact angle changes. We extract the $V_{\text{eff}}$ and $V'_{\text{eff}}$ values at the interface (2950 μm) and plot them in Fig. 4(c). For $V_{\text{eff}}$, they show a constant value at 0.91 V. For $V'_{\text{eff}}$, the plot can be fitted by

$$V'_{\text{eff}} = 0.15 \ln(\kappa) + 0.46. \quad (7)$$

It can be derived that the intersection point is at $\kappa \approx 23.7$. This shows that once the OFF electrode number to ON electrode number ratio is greater than 24, the floating top-electrode device's performance is the same as the grounded top-electrode one. Fig. 4(d) shows the top-electrode potential in the floating top-electrode simulation group and the voltage drop differences at the interface points ($\Delta V_{\text{eff}}$) between the two groups. The two lines have a great consistency over all measured points, and it also can be calculated that both floating top-electrode potential and $\Delta V_{\text{eff}}$ are close to zero once $\kappa \geqq 24$. In addition, since the contact angle profile of the droplet is correlated to the effective voltage in double-plate EWOD system, it can be predicted that the droplet profile in the floating electrode configuration is similar to that in the grounded top-electrode device.

### 4.2 Single droplet movement

We detected the minimum intervals to realize a single droplet continuous moving among linearly arranged electrodes. In general, a small electrode array ($n_1$) requires less time to complete a reliable movement than a big one ($n_2$) regardless of the state of the top-electrode (i.e. floating or grounded).

Fig. 5(b) shows the comparison between grounded top-electrode and floating top-electrode devices with different step intervals. Each point indicates one speed measurement. For $n_1$, the experimental results show that the minimum step interval for 100% completing of a droplet movement over five electrodes is 0.24 s when the top-electrode is grounded. With the same interval, the completion rate for the floating top-electrode setup is only 80% (droplet successfully moves on four out of five electrodes). And a 100% completion can be achieved when increasing the step interval to 0.26 s. Similarly, a 100% completion can be achieved with $n_2$ at 0.31 s (with grounded top-electrode) and 0.32 s (with floating top-electrode). Linear fitting results for the measurements are shown in the sub-table in Fig. 5(b). For both $n_1$ and $n_2$, the grounded top-electrode groups and the floating top-electrode groups show a similar slope, while the floating top-plate lines slightly right shift. In our experiments, the time difference is around 0.01–0.02 s in each step. This is neglectable for sample handling in practical bio-applications since most standard sample preparation will take tens of minutes. Fig. 5(c) shows the snap pictures from the experimental recording video when step interval is 0.22 s, 0.24 s, 0.26 s and 0.31 s. The full video can be found in the ESI.† It worth to mention that there is no additional materials or fabrication steps involved in the floating top-electrode configuration. Compared to conventional grounded top-electrode EWOD system, the only variable affects the droplet moving speed is the effective voltage drop ($V_{\text{eff}}$).

## 5 Conclusions

We report a novel device configuration for double-plate EWOD devices with a floating top-electrode. This can provide the possibilities to integrate additional electrical or electrochemical sensing functions on the top-plate. We carried out both theoretical studies and validation experiments. By increasing the effective grounded electrode area on the bottom-plate, the floating-top-plate setup can perform as good as a grounded top-electrode EWOD device.

## Conflicts of interest

One patent based on this research has been submitted. Y. S. is a co-founder of ACXEL Tech Ltd, and a visiting associate professor in Suzhou Institute of Biomedical Engineering and Technology, Chinese Academy of Sciences. H. M. is a professor in Suzhou Institute of Biomedical Engineering and Technology, Chinese Academy of Sciences, and a co-founder of ACXEL Tech Ltd.

## Acknowledgements

We thank Dr Chen Jiang from University of Cambridge for useful discussion on electro-wetting induced contact angle changes and COMSOL simulations. We thank the technical support from ACXEL's engineering team. This work supported by the National Natural Science Foundation of China (Grant No. 61701493), Policy Guidance project (International Science and Technology Cooperation) of Jiangsu Province of China (BZ2018040), project funded by China Postdoctoral Science Foundation (2019M651959), Postdoctoral Research Funding Program of Jiangsu Province (2018K004B).

## Notes and references

1 M. G. Pollack, R. B. Fair and A. D. Shenderov, *Appl. Phys. Lett.*, 2000, **77**, 1725–1726.
2 J. Lee, H. Moon, J. Fowler, T. Schoellhammer and C.-J. Kim, *Sens. Actuators, A*, 2002, **95**, 259–268.
3 F. Mugele and J.-C. Baret, *J. Phys.: Condens. Matter*, 2005, **17**, R705.
4 R. B. Fair, *Microfluid. Nanofluid.*, 2007, **3**, 245–281.
5 A. R. Wheeler, *Science*, 2008, **322**, 539–540.
6 R. Fobel, C. Fobel and A. R. Wheeler, *Appl. Phys. Lett.*, 2013, **102**, 193513.
7 *Opendrop platform*, http://www.gaudi.ch/OpenDrop/, accessed September 2019.
8 Y.-Y. Lin, R. D. Evans, E. Welch, B.-N. Hsu, A. C. Madison and R. B. Fair, *Sens. Actuators, B*, 2010, **150**, 465–470.
9 A. Rival, D. Jary, C. Delattre, Y. Fouillet, G. Castellan, A. Bellemin-Comte and X. Gidrol, *Lab Chip*, 2014, **14**, 3739–3749.
10 J. A. Schwartz, J. V. Vykoukal and P. R. Gascoyne, *Lab Chip*, 2004, **4**, 11–17.
11 J. H. Noh, J. Noh, E. Kreit, J. Heikenfeld and P. D. Rack, *Lab Chip*, 2012, **12**, 353–360.
12 B. Hadwen, G. Broder, D. Morganti, A. Jacobs, C. Brown, J. Hector, Y. Kubota and H. Morgan, *Lab Chip*, 2012, **12**, 3305–3313.
13 S.-Y. Wu and W. Hsu, *Lab Chip*, 2014, **14**, 3101–3109.
14 C. Peng, Z. Zhang and Y. S. Ju, *Lab Chip*, 2014, **14**, 1117–1122.
15 I. Swyer, S. von der Ecken, B. Wu, A. Jenne, R. Soong, F. Vincent, D. Schmidig, T. Frei, F. Busse and H. J. Stronks, *Lab Chip*, 2019, **19**, 641–653.
16 C. Zhang, Y. Su, S. Hu, K. Jin, Y. Jie, W. Li, A. Nathan and H. Ma, arXiv preprint, arXiv:1909.13085, 2019.
17 S.-K. Fan, T.-H. Hsieh and D.-Y. Lin, *Lab Chip*, 2009, **9**, 1236–1242.
18 C. Jiang, H. Ma, D. G. Hasko and A. Nathan, *Appl. Phys. Lett.*, 2016, **109**, 211601.
19 H. Ren, R. B. Fair, M. G. Pollack and E. J. Shaughnessy, *Sens. Actuators, B*, 2002, **87**, 201–206.
20 R. Bavière, J. Boutet and Y. Fouillet, *Microfluid. Nanofluid.*, 2008, **4**, 287–294.

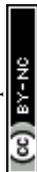